\documentclass[twocolumn,english,prl]{revtex4}
\usepackage[]{fontenc}
\usepackage[latin1]{inputenc}
\usepackage{amsmath}
\usepackage{amssymb}

\makeatletter
\usepackage{amssymb}
\usepackage{babel}

\usepackage{babel}
\makeatother
\begin{document}

\title{Thermal conductivity of a granular metal }

\author{V. Tripathi}

\affiliation{Theory of Condensed Matter Group, Cavendish Laboratory, Department
of Physics, University of Cambridge, Madingley Road, Cambridge CB3
0HE, United Kingdom}

\author{Y. L. Loh}

\affiliation{Theory of Condensed Matter Group, Cavendish Laboratory, Department
of Physics, University of Cambridge, Madingley Road, Cambridge CB3
0HE, United Kingdom}

\date{\today{}}

\begin{abstract}
Using the Kubo formula approach, we study the effect of electron interaction
on thermal transport in the vicinity of a metal-insulator transition,
with a granular metal as our model. For small values of dimensionless
intergrain tunneling conductance, $g\ll1,$ we find that the thermal
conductivity surprisingly shows a phonon-like algebraic decrease,
$\kappa(T)\sim g^{2}T^{3}/E_{c}^{2}$ even though the electrical conductivity
obeys an Arrhenius law, $\sigma(T)\sim ge^{-E_{c}/T};$ therefore
the Wiedemann-Franz (WF) law is seriously violated. We explicitly
show that this violation arises from non-magnetic bosonic excitations
of low energy that transport heat but not charge. At large values
of intergrain tunneling, we find it plausible that the WF law weakly
deviates from the free-electron theory due to potential fluctuations.
Implications for experiment are discussed. 
\end{abstract}
\maketitle
At low temperatures, transport in many metals is dominated by elastic
scattering of electrons. In the absence of phase transitions (such
as superconductivity) and collective modes, 
the Wiedemann-Franz
(WF) law \cite{wiedemann1853} relates the electrical ($\sigma$)
and thermal ($\kappa$) conductivities through a universal Lorenz
number, $L_{0}=\frac{\kappa(T)}{T\sigma(T)}$ where 
$L_{0}\equiv\frac{\pi^{2}k_{B}^{2}}{3e^{2}}$
is Sommerfeld's value for the Lorenz number. Deviations from the 
Wiedemann-Franz
(WF) law beyond those due to phonons and inelastic processes such
as electron-phonon scattering indicate a presence of collective modes
that discriminate between charge and energy flow. Such deviations
are expected to get more pronounced as one approaches a metal-insulator
transition.

The issue of excess electronic thermal conductivity in disordered
metals with interacting electrons has been a topic of much theoretical
debate in recent years. The WF law has been affirmed for any strength
of scattering \cite{chester1961}, disorder \cite{kearney1988}, or
magnetic field \cite{smrcka1977}. In Refs. \cite{langer62,castellani88},
the WF law was shown to hold on account of a Ward identity. Subsequently
there have been many studies including the effect of electron interaction,
some of which have suggested deviations from the WF law 
\cite{livanov1991,arfi1992,niven2005,catelani05,raimondi04,beloborodovheat05}. 

In this paper we study the effect of Coulomb blockade on the thermal
conductivity of a nonmagnetic granular metal. Granular metals are
a novel playground, experimentally and theoretically, for studying
electron interaction effects in the vicinity of a metal-insulator
transition. At not too low temperatures \cite{beloborodov01} transport
in a granular metal is due to a competition of incoherent intergrain
tunneling and Coulomb blockade \cite{ambegaokar82,beloborodov01}.
The resulting model captures the physics of coupling of the continuum
of electron states in the metallic grains with electromagnetic fluctuations.
Granular metals differ from Hubbard models in having a near-continuum
of excitations in each metallic grain. Their study is relevant beyond
artificially-prepared granular metals \cite{gerber97,simon87}. For
example, in more insulating regimes, these materials show a similar
electrical conductivity to disordered semiconductors in the VRH regime.
In more conducting regimes, granular metal physics appears to describe
transport in underdoped cuprate superconductors \cite{ando1,boebinger1}
where there is electronic phase separation into metallic and insulating
regions \cite{lang1}.

We summarize our main results. We obtain for the first time an explicit
Kubo formula for the thermal conductivity of the AES model through
a double functional differentiation of the partition function with
respect to an appropriate source field \cite{luttinger64}. For small values of the dimensionless intergrain conductance, $g\ll1,$ we identify low-energy non-magnetic
bosonic excitations that contribute to heat transport but not charge
transport. Consequently, as our calculations show, the thermal conductivity
decreases only algebraically, $\kappa(T)\sim g^{2}T^{3}/E_{c}^{2},$
quite unlike the electrical conductivity of a regular granular array, which obeys an Arrhenius law, $\sigma(T)\sim ge^{-E_{c}/T}.$ These
bosonic modes are the nearly-continuous particle-hole excitations
in the metallic grains and have no counterpart in a half-filled Hubbard
model with $U\sim E_{c}$ where particle-hole excitations are gapped.
Since the electrical conductivity is insensitive to these neutral
excitations, the algebraic thermal conductivity we calculate distinguishes
between a granular metal and a Hubbard model. This finding also reminds
us of the excess heat transport through excitons proposed recently
\cite{berkovits1999,berkshklovs1999} in the context of disordered
semiconductors. For strong tunneling, $g\gg1,$ we find it plausible that the Lorenz number weakly deviates
from the Sommerfeld value $L_{0}$ due to Coulomb interaction. 

Consider the following microscopic Hamiltonian for a granular metal
array ignoring spin interactions, 
\begin{eqnarray}
H & = & \sum_{\mathbf{x}\lambda}
\bigg[\psi_{\mathbf{x}\lambda}^{\dagger}\xi_{\mathbf{x}
\lambda}\psi_{\mathbf{x}\lambda}+E_{c}\sum_{\mathbf{x}}
(\hat{Q}_{\mathbf{x}}-Q_{0})^{2}\bigg]
(1+h_{\mathbf{x}})+\nonumber \\
+ &  & \frac{1}{2}\sum_{\langle\mathbf{xx}'\rangle,
\lambda\lambda'}(t_{\mathbf{xx}'}^{\lambda\lambda'}
\psi_{\mathbf{x}\lambda}^{\dagger}\psi_{\mathbf{x}'\lambda'}
+\text{h.c.})(1+h_{\mathbf{x}}),
\label{model2}
\end{eqnarray}
where $\textbf{x}$ labels the grains and $\lambda$ labels the states
within each grain, 
$\xi_{\mathbf{x}\lambda}=
\epsilon_{\mathbf{x}\lambda}-\mu.$
$\hat{Q}_{\mathbf{x}}=\sum_{\lambda}
\psi_{\mathbf{x}\lambda}^{\dagger}\psi_{\mathbf{x}\lambda}$
is the charge on grain $\text{\textbf{x},}$ $E_{c}$ is the grain
charging energy, and $t_{\mathbf{xx}'}^{\lambda\lambda'}$ is the
matrix element for nearest-neighbor hopping. Small differences in
temperature of different grains are modeled through the introduction
of {}``gravitational'' potentials $h_{\mathbf{x}}$ \cite{luttinger64}
that couple to the grain electron energy. If $\beta=1/T$ is the average
inverse temperature, then $h_{\mathbf{x}}=(\beta_{\mathbf{x}}-\beta)$
is the small excess of $\beta$ on the $\mathbf{x}^{th}$ grain. 

The strong Coulomb interaction in Eq.(\ref{model2})
makes it difficult to calculate correlation functions directly using the electron
operators. We therefore decouple the interaction to obtain an action for a Hubbard-Stratonovich field, $V$, representing the electrostatic potential on each grain,
\begin{eqnarray}
& &S_{0}[V] = -\frac{1}{4E_{c}}\sum_{\mathbf{x}}\int_{0}^{\beta}d\tau\,
\frac{V_{\tau\mathbf{x}}^{2}}{1+h_{\tau\mathbf{x}}}+\nonumber \\
& &+ \sum_{\mathbf{x}\mathbf{x}';\lambda\lambda'}
\int_{0}^{\beta}d\tau\, t_{\mathbf{x}\mathbf{x}'}^{\lambda\lambda'}
\psi_{\mathbf{x}\lambda}^{\dagger}\psi_{\mathbf{x}'\lambda'}
(1+\frac{h_{\tau\mathbf{x}}+h_{\tau\mathbf{x}'}}{2})+\nonumber \\
& &+ \sum_{\mathbf{x}\lambda} \int_{0}^{\beta}d\tau\,
\psi_{\mathbf{x}\lambda}^{\dagger}(\partial_{\tau}+
V_{\tau\mathbf{x}}+\xi_{\mathbf{x}\lambda}
(1+h_{\tau\mathbf{x}}))\psi_{\mathbf{x}
\lambda},
\label{HuSt}
\end{eqnarray}
where $\tau$ is imaginary time, and the $h_{\mathbf{x}}$ 
sources are now $\tau-$dependent.
We then make the gauge transformation $\psi_{\mathbf{x}\lambda}\rightarrow
e^{-i\varphi_{\tau\mathbf{x}}}\psi_{\mathbf{x}\lambda}$ and 
$\, V_{\mathbf{x}}=i\partial_{\tau}\varphi_{\tau\mathbf{x}}$.  This eliminates the $\psi^\dag V \psi$ term, at the cost of replacing $t$ by a gauge-dependent hopping amplitude,
$\tilde{t}_{\tau\mathbf{xx'}}^{\lambda\lambda'} 
 =  t_{\mathbf{xx}'}^{\lambda\lambda'}
e^{i\varphi_{\tau\mathbf{xx}'}},$ where 
$
\varphi_{\tau\mathbf{xx}'} =
\varphi_{\tau\mathbf{x}}-\varphi_{\tau\mathbf{x}'}
$.
The Matsubara fields $\varphi_{\tau\mathbf{x}}$ satisfy bosonic
boundary conditions, 
$\varphi_{\beta\mathbf{x}}=2\pi k_{\mathbf{x}}+\varphi_{0\mathbf{x}},$
where $k_{\mathbf{x}}\in\mathbb{Z}$ is the winding number at site
$\mathbf{x}.$ 

Next we integrate out the fermions and expand the electron
determinant to $O(h\tilde{t}^2)$,
\begin{eqnarray}
S[\varphi,h]\approx\frac{1}{4E_{c}}
\sum_{\mathbf{x}}\int_{\tau}
\frac{(\partial_{\tau}\varphi_{\tau\mathbf{x}})^{2}}
{1+h_{\tau\mathbf{x}}}+\text{tr}
[G_{\mathbf{x}\lambda}(\tau,\tau)h_{\tau\mathbf{x}}
\xi_{\mathbf{x}\lambda}]+\nonumber \\
+\frac{1}{2}\text{tr}
[G_{\mathbf{x}\lambda}(\tau_{1},\tau_{2})
\tilde{t}_{\tau_{2}\mathbf{xx}'}^{\lambda\lambda'}
G_{\mathbf{x}'\lambda'}(\tau_{2},\tau_{1})
\tilde{t}_{\tau_{1}\mathbf{x}'\mathbf{x}}^{\lambda'\lambda}]
+\nonumber \\
+\frac{1}{2}\text{tr}[G_{\mathbf{x}\lambda}(\tau_{1},\tau_{2})
\tilde{t}_{\tau_{2}\mathbf{xx}'}^{\lambda\lambda'}
G_{\mathbf{x}'\lambda'}(\tau_{2},\tau_{1})
\tilde{t}_{\tau_{1}\mathbf{x}'\mathbf{x}}^{\lambda'\lambda}
(h_{\tau_{2}\mathbf{x}}+h_{\tau_{2}\mathbf{x}'})]
+\nonumber \\
+\text{tr}[G_{\mathbf{x}\lambda}(\tau_{1},\tau_{2})
h_{\tau_{2}\mathbf{x}}\xi_{\mathbf{x}\lambda}
G_{\mathbf{x}\lambda}(\tau_{2},\tau_{3})
\tilde{t}_{\tau_{3}\mathbf{xx}'}^{\lambda\lambda'}
G_{\mathbf{x}'\lambda'}(\tau_{3},\tau_{1})
\tilde{t}_{\tau_{1}\mathbf{x}'\mathbf{x}}^{\lambda'\lambda}],
\label{action1}
\end{eqnarray}
where ``tr'' means a sum over all indices, 
and $G_{\mathbf{x}\lambda}(\tau,\tau')$ is the 
electron Green function
\begin{equation}
G_{\mathbf{x}\lambda}(\tau,\tau') = T\sum_{n}\frac{e^{-i\nu_{n}(\tau-\tau')}}
{i\nu_{n}-\xi_{\mathbf{x}\lambda}}
= [n(\xi_{\mathbf{x}\lambda})-\Theta_{\tau-\tau'}]
e^{-\xi_{\mathbf{x}\lambda}(\tau-\tau')}.
\label{greenfn}
\end{equation}
Here $\nu_{n}=2\pi T(n+1/2)$ and $\Theta_{\tau}$ is the unit step
function. We will also need the momentum-summed Green function, 
$G(\tau,\tau')=\sum_{\lambda}G_{\mathbf{x}\lambda}(\tau,\tau')$,
which is independent of the grain label. If the temperature is much
larger than the Thouless energy for intergrain diffusion, 
$T\gg|t|^{2}\nu(\epsilon_{F}),$
and if the grains are much larger than the Fermi wavelength of the
metal, it suffices to expand the electron determinant to second order
in tunneling \cite{zarand2000}.

The (time-dependent) energy of the $\mathbf{x}^{th}$ grain, 
$E_{\tau\mathbf{x}},$
is obtained by differentiating the action in Eq.(\ref{action1}) with
respect to $h_{\tau\mathbf{x}}.$ Dropping any constant terms, and
using the above Green functions, we have, 
\begin{eqnarray}
E_{\tau\mathbf{x}} & = & -\frac{1}{4E_{c}}
(\partial_{\tau}\varphi_{\tau\mathbf{x}})^{2}+\sum_{\mathbf{x}'
=\mathbf{x}\pm\mathbf{a}}|t|^{2}
\bigg[\int_{\tau_{1}}G(\tau_{1},\tau)\times\nonumber \\
\times &  & \!\!\!\!\!\!\!\!\!\!\! G(\tau,\tau_{1})
\cos(\varphi_{\tau\mathbf{xx'}}-\varphi_{\tau_{1}\mathbf{xx'}})
+\sum_{\lambda\lambda'}\!\!\!\int_{\tau_{1}\tau_{2}}\!\!\!\!
\xi_{\mathbf{x}\lambda}G_{\mathbf{x}\lambda}(\tau_{1},\tau)
\times\nonumber \\
\times &  & G_{\mathbf{x}\lambda}(\tau,\tau_{2})
G_{\mathbf{x'}\lambda'}(\tau_{2},\tau_{1})
\exp[i(\varphi_{\tau_{2}\mathbf{xx'}}-
\varphi_{\tau_{1}\mathbf{xx'}})]\bigg],
\label{energy1}
\end{eqnarray}
where we assumed that the bare tunneling element is a constant. To
obtain an expression for the energy current $J_{\tau\mathbf{xx'}}$
\emph{from grain $\mathbf{x}$ to a neighboring grain $\mathbf{x'},$}
we employ the continuity equation, 
$\partial_{\tau}E_{\tau\mathbf{x}}=i\sum_{\mathbf{x'}}J_{\tau\mathbf{xx'}}.$
We will need the equation of motion for $\varphi_{\tau\mathbf{x}}$
from Eq.(\ref{action1}) (with sources set to zero) and the Green
functions defined in Eq.(\ref{greenfn}). 
The symmetrized energy current, 
$j_{\tau\mathbf{xx'}}^{(E)}=\frac{1}{2} 
(J_{\tau\mathbf{xx'}}-J_{\tau\mathbf{x'x}})$, is
\begin{eqnarray}
j_{\tau\mathbf{xx'}}^{(E)} & = & i\pi gT^{2}
\int_{\tau'}\frac{1}{2}(\partial_{\tau}\varphi_{\tau\mathbf{x}}
+\partial_{\tau}\varphi_{\tau\mathbf{x'}})
\frac{1}{\sin^{2}[\pi T(\tau-\tau')]}\times\nonumber \\
\times &  & \!\!\!\!\!\!\!\!\!\!
\sin(\varphi_{\tau\mathbf{xx'}}-\varphi_{\tau'\mathbf{xx'}})
-\pi gT^{2}\int_{\tau'}\bigg[\partial_{\tau}
\frac{1}{\sin[\pi T(\tau-\tau')]}\bigg]
\times\nonumber \\
\times &  & \!\!\!\!\!\!\!\!\!\!\frac{1}{\sin[\pi T(\tau-\tau')]}
\sin(\varphi_{\tau\mathbf{xx'}}-\varphi_{\tau'\mathbf{xx'}}),
\label{thermalcurrent3}
\end{eqnarray}
where we have introduced the dimensionless intergrain tunneling conductance
$g=2\pi\nu^{2}(\epsilon_{F})|t|^{2}$ and used 
$G(\tau,\tau')=\frac{\pi\nu(\epsilon_{F})T}{\sin \pi T(\tau-\tau')}.$
To obtain the Kubo formula for thermal conductivity, we need to 
calculate the non-local specific heat through a double differentiation
of the partition function with respect to $h_{\tau\mathbf{x}},$ and
then extract the thermal conductivity using the continuity equation.

The same result may be obtained more easily by introducing a source
field $f_{\tau\mathbf{x}}$ that can generate the energy current,
$h_{\tau\mathbf{x}}=i\partial_{\tau}f_{\tau\mathbf{x}}.$ In Eq.(\ref{HuSt})
we expand to linear order in $h_{\tau\mathbf{x}}$ and use the field
$f_{\tau\mathbf{x}}$ instead of $h_{\tau\mathbf{x}}.$ 
Next we integrate out the conduction electrons and expand 
to second order in tunneling. 
The resulting action is
\begin{eqnarray}
S[\varphi,f]\approx\frac{1}{4E_{c}}
\sum_{\mathbf{x}}\int_{\tau}
(\partial_{\tau}\varphi_{\tau\mathbf{x}})^{2}+
\qquad\qquad\qquad\qquad\nonumber \\
+\frac{1}{2}\text{tr}[G_{\mathbf{x}\lambda}(\tau_{1},\tau_{2})
\hat{t}_{\tau_{2}\mathbf{xx}'}^{\lambda\lambda'}
G_{\mathbf{x}'\lambda'}(\tau_{2},\tau_{1})
\hat{t}_{\tau_{1}\mathbf{x}'\mathbf{x}}^{\lambda'\lambda}],
\label{action2}\\
\hat{t}_{\tau\mathbf{xx}'}^{\lambda\lambda'}=
\tilde{t}_{\tau\mathbf{xx'}}^{\lambda\lambda'}
\left[
1+if_{\tau\mathbf{xx}'}
\left((\xi_{\mathbf{x}\lambda}+\xi_{\mathbf{x}'\lambda'})/2+
V^{(av)}_{\tau\mathbf{x}\mathbf{x}'}\right)
\right],
\label{tunnelmod}
\end{eqnarray}
where $f_{\tau\mathbf{xx}'}=
f_{\tau\mathbf{x}}-f_{\tau\mathbf{x}'}$ and 
$V^{(av)}_{\tau\mathbf{x}\mathbf{x}'}=
(V_{\tau\mathbf{x}}+V_{\tau\mathbf{x}'})/2$. 
The energy current in Eq.(\ref{thermalcurrent3}) 
can be obtained from Eq.(\ref{action2})
using $j_{\tau\mathbf{xx'}}^{(E)}
[\varphi,f]=\delta S[\varphi,f]/\delta f_{\tau\mathbf{xx'}}.$
This is similar to the derivation of the electric current,
with electric charge replaced by electronic energy.
The thermal conductivity $\kappa$
may be obtained by further differentiation, 
\begin{eqnarray}
\kappa(\omega,T) & = & 
ia^{2-d}\frac{1}{\omega}\int_{0}^{\beta}d\tau\, 
e^{i\Omega_{n}\tau}K^{(E)}(\tau)\bigg|_{\Omega_{n}\rightarrow-i\omega^{+}}\!\!\!\!\!\!\!\!\!\!\!\!\!\!\!\!\!\!\!,
\label{thermconddef}
\\
K^{(E)}(\tau_{1}-\tau_{2}) & = & 
\!\!\!\frac{\beta}{Z}\int\!\!\! 
D\varphi\, e^{-S}\left[
\frac{\delta j_{\tau_{2}\mathbf{x},\mathbf{x}+\mathbf{a}}^{(E)}
[\varphi,f]}{\delta f_{\tau_{1}\mathbf{x},\mathbf{x}+\mathbf{a}}}
\right.-
\nonumber \\
 & - & \left.\sum_{\mathbf{x}'}\!\!\int\!\!\, 
j_{\tau_{2}\mathbf{x}',\mathbf{x}'+\mathbf{a}}^{(E)}
\frac{\delta S[\varphi,f]}{\delta f_{\tau_{1}\mathbf{x}',
\mathbf{x}'+\mathbf{a}}}\right]\bigg|_{f=0}\!\!\!\!\!,
\label{kappadef}
\end{eqnarray}
where $\omega^{+}=\omega+i\delta.$ The first term, which has come
to be known as the {}``diamagnetic'' contribution $\kappa^{(d)}$,
is a local term in the sense that it is an average on a single bond.
The second term in Eq.(\ref{kappadef}) is a product of two energy
currents, and is also known as the {}``paramagnetic'' contribution
$\kappa^{(p)}$. The diamagnetic contribution 
may be obtained by differentiating 
Eq.(\ref{action2}) twice with respect to the source, 
\begin{eqnarray}
& & \kappa^{(d)}(\omega,T) = 
-ia^{2-d}\frac{\pi^{2}gT^{3}}{\omega}\int_{0}^{\beta}
d\tau\,(e^{i\Omega_{n}\tau}-1)\times\nonumber \\
& &
\left[\tfrac{1}{\sin \pi T\tau}\tfrac{\partial^{2}}
{\partial_{\tau_{c}}^{2}}\tfrac{1}
{\sin \pi T(\tau+i\tau_{c})}-
V_{\tau\mathbf{x},\mathbf{x}+\mathbf{a}}^{(av)}
V_{0\mathbf{x},\mathbf{x}+\mathbf{a}}^{(av)}
\tfrac{1}{\sin^{2} \pi T \tau}
\right]
\nonumber \\
 &  & \times\left\langle 
\cos\left(\varphi_{\tau \mathbf{x},\mathbf{x}+\mathbf{a}}-
\varphi_{0\mathbf{x},\mathbf{x}+\mathbf{a}}\right)\right
\rangle \bigg|_{\Omega_{n}\rightarrow-i\omega^{+},\tau_{c}\rightarrow 0}.
\label{kappadia2}
\end{eqnarray}
Let us call the $V$-independent part of Eq.(\ref{kappadia2}) the
kinetic contribution, $\kappa^{(d,K)}$, and the $V$-dependent part
the potential contribution $\kappa^{(d,V)}.$ 
Eq.(\ref{kappadia2}) is similar to the known expression for the diamagnetic
contribution to the electrical conductivity obtained in Ref.\cite{efetov02}.

\textbf{A.} Consider first weakly coupled grains, $g\ll1.$ 
Do a perturbation expansion in $g$,
$\kappa=\kappa^{(1)}+\kappa^{(2)}+\cdots,$ where $\kappa^{(n)} \sim g^n$.
The diamagnetic parts of the electrical \cite{efetov02} and thermal
conductivities [Eq.(\ref{kappadia2}] involve the bond correlator,
$\Pi_{\tau}=\langle
\exp i(\varphi_{\tau\mathbf{x},\mathbf{x}
+\mathbf{a}}-\varphi_{0\mathbf{x},\mathbf{x}+\mathbf{a}})
\rangle=\Pi^{(0)}+\Pi^{(1)}+\cdots.$
Because of the Coulomb blockade,
$\Pi_{\tau}^{(0)}  =  \frac{1}{Z}
\sum_{q_{1},q_{2}}e^{-(q_{1}^{2}+q_{2}^{2})
\beta E_{c}-2E_{c}\tau(1-q_{1}-q_{2})}$
is exponential in $\tau$.
So
\begin{align}
\sigma^{(d,1)} &= ge^{2}a^{2-d}
(2e^{-\beta E_{c}}+2\beta E_{c}e^{-2\beta E_{c}}),\nonumber \\
\kappa^{(d,K,1)} &= gTa^{2-d} 
(\tfrac{2\pi^{2}}{3}e^{-\beta E_{c}}+
\tfrac{8}{3}(\beta E_{c})^{3}e^{-2\beta E_{c}}).
\label{Ogthermaldia}
\end{align}
In contrast, $\Pi_{\tau}^{(1)}$  has a 
\emph{power-law} $\tau$-dependence \cite{lohgranular05}, 
which is a result of the $1/\tau^2$ interaction in the AES action, 
Eq.\eqref{action2}, arising from 
\emph{particle-hole inelastic cotunneling}:
\begin{eqnarray*}
\Pi_{\tau}^{(1)} = \frac{2\pi gT^{2}}{E_{c}^{2}
\sin^{2}(\pi T\tau)}=\frac{2\pi g}{E_{c}^{2}}\alpha_{\tau}, \quad
\alpha_{\tau}=\frac{T^{2}}{\sin^{2}(\pi T\tau)}.
\end{eqnarray*}
It gives rise to $O(g^2)$ terms in both the diamagnetic and paramagnetic contributions to $\sigma$ and $\kappa$. 
We have shown \cite{lohgranular05} in a previous work that
\begin{align*}
\sigma^{(d,2)}\!\!\! & =
\!\!-\sigma^{(p,2)}\!\!\! 
=\!\!\frac{i\pi ga^{2-d}e^{2}}{\omega}
\!\!\left(\frac{2\pi g}{E_{c}^{2}}\right)\!\!\!\int_{\tau}
(e^{i\Omega_{n}\tau}-1)\alpha_{\tau}^{2}
\bigg|_{\Omega_{n}\rightarrow-i\omega^{+}}
\!\!\!
\\&\approx
\frac{4\pi}{3}e^{2}a^{2-d}\left(\frac{gT}{E_{c}}\right)^{2};
\end{align*}
the power-law terms in the electrical conductivity 
\emph{cancel} \cite{lohgranular05},
leaving an Arrhenius law $\sigma=\sigma^{(d,1)}\sim e^{-E_c/T}.$ 
Now consider the $O(g^{2})$
terms in the thermal conductivity. The dominant contributions to both
the diamagnetic and paramagnetic terms arise from charge-neutral configurations
($q_{1}=q_{2}=0$), and, unlike the $O(g)$ terms, are not Arrhenius-suppressed:
\begin{align}
\kappa^{(d,K,2)} & =  \frac{i\pi ga^{2-d}}{\omega}
\!\!\left(\frac{2\pi g}{TE_{c}^{2}}\right)
\!\!\int_{\tau}(e^{i\Omega_{n}\tau}-1)
\gamma_{\tau}\alpha_{\tau}\bigg|_{\Omega_{n}\rightarrow-i\omega^{+}},
\nonumber \\
\kappa^{(p,K,2)} & =  
\frac{i\pi ga^{2-d}}{4\omega}\!\!
\left(\frac{2\pi g}{TE_{c}^{2}}\right)\!\!
\int_{\tau}(e^{i\Omega_{n}\tau}-1)\alpha_{\tau}'
\alpha_{\tau}'\bigg|_{\Omega_{n}\rightarrow-i\omega^{+}},
\label{Og2thermal}
\end{align}
where $\gamma_{\tau}=T^{2}(\sin(\pi T\tau))^{-1}
\partial_{\tau}^{2}(\sin(\pi T\tau)^{-1}.$
These two terms have the same sign and \emph{do not cancel},
\begin{eqnarray}
\kappa^{(d,K,2)} 
=3\kappa^{(p,K,2)}
= \frac{12\pi^{3}}{15}a^{2-d}
\left(\frac{g^{2}T^{3}}{E_{c} {}^{2}}\right),
\label{noncancellation}
\end{eqnarray}
and so the thermal
conductivity is only algebraically small,
\begin{eqnarray}
\kappa^{(K,2)} & = & (16\pi^{3}/15)
(g^{2}T^{3}/E_{c} {}^{2}).
\label{centralresult}
\end{eqnarray}
\emph{This is the central result of our paper.}
In the argument above, the algebraic temperature dependence 
is an eventual result of the 
\emph{particle-hole inelastic cotunneling processes} 
described by the AES action.

The potential energy contributions 
$\kappa^{(V,2)}$ and $\kappa^{(V,2)}$, which involve
averages such as
$\langle V_{\tau\mathbf{x},\mathbf{x}+\mathbf{a}}^{(av)}
V_{0\mathbf{x},\mathbf{x}+\mathbf{a}}^{(av)}
\cos(\varphi_{\tau\mathbf{x},\mathbf{x}+\mathbf{a}}
-\varphi_{0\mathbf{x},\mathbf{x}+\mathbf{a}})
\rangle$,
 turn out to be proportional to $(q_{1}-q_{2})^{2}.$
This immediately rules out neutral processes,
such as inelastic cotunneling, for which $q_{1}=q_{2}=0.$
The $O(g^0)$ contribution to the above average can be shown
to be 
$E_{c}^{2}\sum_{q_{1},q_{2}}(q_{1}-q_{2})^{2}
e^{-(q_{1}^{2}+q_{2}^{2})\beta E_{c}-2E_{c}\tau(1-q_{1}-q_{2})}$, so that 
$\kappa^{(V,2)} \sim O(e^{-\beta E_{c}}),$ which can be neglected in comparison with
$\kappa^{(K,2)}$.

\textbf{B.} Consider now the case where the dimensionless intergrain
tunneling conductance $g$ is large, $g=\pi|t|^{2}\nu(\epsilon_{F})^{2}.$
One can show \cite{efetov02} 
that the diamagnetic contribution to the electrical conductivity is 
$\sigma^{(d)} = 
e^{2}a^{2-d}g\left(1-\frac{1}{\pi gz}\ln \frac{gE_{c}}{T}\right).$
Comparing with
Eq.(\ref{kappadia2}),
it is evident that for the diamagnetic components, the
kinetic part of the thermal conductivity is simply related to the
electrical conductivity by
$\kappa^{(d,K)}/(T\sigma^{(d)})=\frac{\pi^{2}k_{B}^{2}}{3e^{2}}$,
where we have restored the Boltzmann
constant $k_{B}.$ So at this level the WF law is obeyed. Next consider
the kinetic part of the paramagnetic contribution
to the thermal conductivity, $\kappa^{(p,K)}(T)$.  Ref.\cite{efetov02} shows that $\sigma^{(p)} \ll \sigma^{(d)}$; adopting a similar argument, one can show that
 $\kappa^{(p,K)}$ is smaller
than $\kappa^{(d,K)}$ by a factor $g^{-2},$ and furthermore it does
not have a logarithmic singularity. We have not yet considered the
extra contributions from the potential part, $\kappa^{(d,V)}$ and
$\kappa^{(p,V)}.$ Evaluating $\kappa^{(d,V)}$ for example leads
to corrections similar to those obtained in Ref.\cite{beloborodovheat05}.
However since the potential part $\kappa^{(p,V)}$ of the paramagnetic
term is of the order of $\kappa^{(d,V)},$ a proper treatment should
consider both contributions. This could, in principle, lead to a different
conclusion from Ref.\cite{beloborodovheat05}. So at this stage we
are only able to make a weaker statement that for the strong tunneling
case at not too low temperatures ($T\gg gE_{c}e^{-\pi gz}$),
$\frac{\kappa}{T\sigma} =  
\frac{\pi^{2}k_{B}^{2}}{3e^{2}} +O(g^{-1}).$
Thus the Lorenz number $L$ is larger than Sommerfeld's free electron
value by $O(g^{-1})$ corrections.

We conclude with a discussion of our results and comments on existing
and future experiment. We have shown that thermal transport in granular
metals at low temperatures is dominated by cotunneling of low-energy
electron-hole pairs. These neutral excitations do not transport charge.
As a result, while the electrical conductivity for weakly coupled
grains is exponentially small in temperature, 
$\sigma(T)\sim e^{2}ga^{2-d}e^{-E_{c}/T},$
the thermal conductivity is only algebraically small, 
$\kappa(T)\sim a^{2-d}k_{B}^{2}g^{2}T^{3}/E_{c}^{2}.$
The particle-hole cotunneling process is physically equivalent to
a particle cotunneling loop. An electron executing a cotunneling loop
brings back its charge to the starting grain and hence there is little
change in the electrical conductivity. There is no requirement, however,
that the returning electron has exactly the same energy. This conservation
of grain charge but not grain energy in a cotunneling loop is at the
heart of the difference between heat and charge transport. 
The energy gained by an electron due to nearest neighbor
tunneling is of the order of $(t^{2}/E_{c}),$ and the tunneling probability
is, roughly, $(t/E_{c})^{2}.$ The number of particle-hole excitations
in each grain is of the order of $\nu(\epsilon_{F})T\gg1,$ and the
energy of a particle-hole is of the order of $T.$ This gives us a
rough estimate $\kappa(T)\propto g^{2}T^{3}/E_{c}^{2},$ in agreement
with our detailed calculation. 
The near-continuum of states in the
metallic grains means the electron-hole excitations can have very
low energy; in this sense cotunneling in regular granular metals differs
fundamentally from exciton transport in the Hubbard model even when
intergrain tunneling is small. However there seems to be some resonance
with excess heat transport by delocalized tightly-bound excitons in
dirty semiconductors \cite{berkovits1999,berkshklovs1999}. Experimentally,
$Al-Ge$ granular systems in the $g\gg1$ regime \cite{shapira84}
show a linear$-T$ thermal conductivity with the value of the Lorenz
number somewhat larger than $L_{0},$ $L\approx1.8L_{0}$ at the lowest
temperatures measured. This is consistent with our assessment. 

Our predictions can be experimentally verified:
even though the $g^{2}T^{3}/E_{c}^{2}$
law we obtain for $\kappa(T)$ is reminiscent of the $T^{3}$ phonon
contribution, it can be distinguished experimentally through its dependence
on $g$ and $E_{c}.$ Besides, the electronic contribution may be of the order
of the phonon contribution as can be seen in the following rough estimate. 
Let $a$ denote the distance between grain centers,
$l$ the phonon mean free path, $c$ the speed of sound, $n$ the atomic number density,
and $\Theta_D$ the Debye temperature of the material. The ratio of the
electronic and phononic thermal conductivities is then of the order of 
$r = (g \Theta_D /E_c)^2 (k_B \Theta_D)/(\hbar a n c l)$. In a typical
three-dimensional granular metal, the grain diameter is of the order of $100$\AA\ and the insulating
space between the grains is about $10$\AA.  With $\Theta_D \sim 300 K$, $E_c \sim 100 K$, $g\lesssim 1$, $a\sim
100$\AA, $l \sim 10$\AA \footnote{The phonons are assumed to
scatter randomly at the boundaries of the metallic and insulating regions, so
their mean free path in the insulator, $l$, is also of the order of
$10$\AA.}, $c\sim 10^3 m/s$, and $n\sim 10^{28}m^{-3}$, we estimate 
$r \approx g^2 \sim O(1)$. 
The electronic contribution can be further enhanced by increasing the
size of the grains and choosing an insulator with a higher dielectric
constant; each of these decreases $E_c$. Physically, the surprisingly large
electronic thermal conductivity can be attributed to the large
electron speed compared to the speed of sound. 
\begin{acknowledgments}
We are grateful to D. E. Khmelnitskii for valuable discussions. V.T.
thanks Trinity College, Cambridge for a JRF, and Y.L.L. thanks Trinity
College and Cavendish Laboratory, Cambridge for support. 
\end{acknowledgments}

\begin{thebibliography}{26}
\expandafter\ifx\csname natexlab\endcsname\relax\def\natexlab#1{#1}\fi
\expandafter\ifx\csname bibnamefont\endcsname\relax
  \def\bibnamefont#1{#1}\fi
\expandafter\ifx\csname bibfnamefont\endcsname\relax
  \def\bibfnamefont#1{#1}\fi
\expandafter\ifx\csname citenamefont\endcsname\relax
  \def\citenamefont#1{#1}\fi
\expandafter\ifx\csname url\endcsname\relax
  \def\url#1{\texttt{#1}}\fi
\expandafter\ifx\csname urlprefix\endcsname\relax\def\urlprefix{URL }\fi
\providecommand{\bibinfo}[2]{#2}
\providecommand{\eprint}[2][]{\url{#2}}

\bibitem[{\citenamefont{Wiedemann and Franz}(1853)}]{wiedemann1853}
\bibinfo{author}{\bibfnamefont{G.~H.} \bibnamefont{Wiedemann}}
  \bibnamefont{and} \bibinfo{author}{\bibfnamefont{R.}~\bibnamefont{Franz}},
  \bibinfo{journal}{Ann. Phys., Leipzig,(2)} \textbf{\bibinfo{volume}{89}},
  \bibinfo{pages}{497} (\bibinfo{year}{1853}).

\bibitem[{\citenamefont{Chester and Thellung}(1961)}]{chester1961}
\bibinfo{author}{\bibfnamefont{G.~V.} \bibnamefont{Chester}} \bibnamefont{and}
  \bibinfo{author}{\bibfnamefont{A.}~\bibnamefont{Thellung}},
  \bibinfo{journal}{Proc. Phys. Soc.} \textbf{\bibinfo{volume}{77}},
  \bibinfo{pages}{1005} (\bibinfo{year}{1961}).

\bibitem[{\citenamefont{Kearney and Butcher}(1988)}]{kearney1988}
\bibinfo{author}{\bibfnamefont{M.~J.} \bibnamefont{Kearney}} \bibnamefont{and}
  \bibinfo{author}{\bibfnamefont{P.~N.} \bibnamefont{Butcher}},
  \bibinfo{journal}{J. Phys. C: Solid State Phys.}
  \textbf{\bibinfo{volume}{21}}, \bibinfo{pages}{L265} (\bibinfo{year}{1988}).

\bibitem[{\citenamefont{Smrcka and Streda}(1977)}]{smrcka1977}
\bibinfo{author}{\bibfnamefont{L.}~\bibnamefont{Smrcka}} \bibnamefont{and}
  \bibinfo{author}{\bibfnamefont{P.}~\bibnamefont{Streda}},
  \bibinfo{journal}{J. Phys. C: Solid State Phys.}
  \textbf{\bibinfo{volume}{10}}, \bibinfo{pages}{2153} (\bibinfo{year}{1977}).

\bibitem[{\citenamefont{Langer}(1962)}]{langer62}
\bibinfo{author}{\bibfnamefont{J.~S.} \bibnamefont{Langer}},
  \bibinfo{journal}{Phys. Rev.} \textbf{\bibinfo{volume}{128}},
  \bibinfo{pages}{110} (\bibinfo{year}{1962}).

\bibitem[{\citenamefont{Castellani et~al.}(1988)\citenamefont{Castellani,
  DiCastro, Kotliar, Lee, and Strinati}}]{castellani88}
\bibinfo{author}{\bibfnamefont{C.}~\bibnamefont{Castellani}},
  \bibinfo{author}{\bibfnamefont{C.}~\bibnamefont{DiCastro}},
  \bibinfo{author}{\bibfnamefont{G.}~\bibnamefont{Kotliar}},
  \bibinfo{author}{\bibfnamefont{P.~A.} \bibnamefont{Lee}}, \bibnamefont{and}
  \bibinfo{author}{\bibfnamefont{G.}~\bibnamefont{Strinati}},
  \bibinfo{journal}{Phys. Rev. \textbf{B}} \textbf{\bibinfo{volume}{37}},
  \bibinfo{pages}{9046} (\bibinfo{year}{1988}).

\bibitem[{\citenamefont{Livanov et~al.}(1991)\citenamefont{Livanov, Reizer, and
  Sergeev}}]{livanov1991}
\bibinfo{author}{\bibfnamefont{D.~V.} \bibnamefont{Livanov}},
  \bibinfo{author}{\bibfnamefont{M.~Y.} \bibnamefont{Reizer}},
  \bibnamefont{and} \bibinfo{author}{\bibfnamefont{A.~V.}
  \bibnamefont{Sergeev}}, \bibinfo{journal}{JETP}
  \textbf{\bibinfo{volume}{72}}, \bibinfo{pages}{760} (\bibinfo{year}{1991}).

\bibitem[{\citenamefont{Arfi}(1992)}]{arfi1992}
\bibinfo{author}{\bibfnamefont{B.}~\bibnamefont{Arfi}}, \bibinfo{journal}{J.
  Low Temp. Phys.} \textbf{\bibinfo{volume}{86}}, \bibinfo{pages}{213}
  (\bibinfo{year}{1992}).

\bibitem[{\citenamefont{Niven and Smith}(2005)}]{niven2005}
\bibinfo{author}{\bibfnamefont{D.~R.} \bibnamefont{Niven}} \bibnamefont{and}
  \bibinfo{author}{\bibfnamefont{R.~A.} \bibnamefont{Smith}},
  \bibinfo{journal}{Phys. Rev. \textbf{B}} \textbf{\bibinfo{volume}{71}},
  \bibinfo{pages}{035106} (\bibinfo{year}{2005}).

\bibitem[{\citenamefont{Catelani and Aleiner}(2005)}]{catelani05}
\bibinfo{author}{\bibfnamefont{G.}~\bibnamefont{Catelani}} \bibnamefont{and}
  \bibinfo{author}{\bibfnamefont{I.~L.} \bibnamefont{Aleiner}},
  \bibinfo{journal}{JETP} \textbf{\bibinfo{volume}{100}}, \bibinfo{pages}{331}
  (\bibinfo{year}{2005}).

\bibitem[{\citenamefont{Raimondi et~al.}(2004)\citenamefont{Raimondi, Savona,
  Schwab, and L\"uck}}]{raimondi04}
\bibinfo{author}{\bibfnamefont{R.}~\bibnamefont{Raimondi}},
  \bibinfo{author}{\bibfnamefont{G.}~\bibnamefont{Savona}},
  \bibinfo{author}{\bibfnamefont{P.}~\bibnamefont{Schwab}}, \bibnamefont{and}
  \bibinfo{author}{\bibfnamefont{T.}~\bibnamefont{L\"uck}},
  \bibinfo{journal}{Phys. Rev. \textbf{B}} \textbf{\bibinfo{volume}{70}},
  \bibinfo{pages}{155109} (\bibinfo{year}{2004}).

\bibitem[{\citenamefont{Beloborodov et~al.}(2005)\citenamefont{Beloborodov,
  Lopatin, Hekking, Fazio, and Vinokur}}]{beloborodovheat05}
\bibinfo{author}{\bibfnamefont{I.~S.} \bibnamefont{Beloborodov}},
  \bibinfo{author}{\bibfnamefont{A.~V.} \bibnamefont{Lopatin}},
  \bibinfo{author}{\bibfnamefont{F.~W.~J.} \bibnamefont{Hekking}},
  \bibinfo{author}{\bibfnamefont{R.}~\bibnamefont{Fazio}}, \bibnamefont{and}
  \bibinfo{author}{\bibfnamefont{V.~M.} \bibnamefont{Vinokur}},
  \bibinfo{journal}{Europhys. Lett.} \textbf{\bibinfo{volume}{69}},
  \bibinfo{pages}{435} (\bibinfo{year}{2005}).

\bibitem[{\citenamefont{Beloborodov et~al.}(2001)\citenamefont{Beloborodov,
  Efetov, Altland, and Hekking}}]{beloborodov01}
\bibinfo{author}{\bibfnamefont{I.~S.} \bibnamefont{Beloborodov}},
  \bibinfo{author}{\bibfnamefont{K.~B.} \bibnamefont{Efetov}},
  \bibinfo{author}{\bibfnamefont{A.}~\bibnamefont{Altland}}, \bibnamefont{and}
  \bibinfo{author}{\bibfnamefont{F.~W.~J.} \bibnamefont{Hekking}},
  \bibinfo{journal}{Phys. Rev. B} \textbf{\bibinfo{volume}{63}},
  \bibinfo{pages}{115109} (\bibinfo{year}{2001}).

\bibitem[{\citenamefont{Ambegaokar et~al.}(1982)\citenamefont{Ambegaokar,
  Eckern, and Sch{\"o}n}}]{ambegaokar82}
\bibinfo{author}{\bibfnamefont{V.}~\bibnamefont{Ambegaokar}},
  \bibinfo{author}{\bibfnamefont{U.}~\bibnamefont{Eckern}}, \bibnamefont{and}
  \bibinfo{author}{\bibfnamefont{G.}~\bibnamefont{Sch{\"o}n}},
  \bibinfo{journal}{Phys. Rev. Lett.} \textbf{\bibinfo{volume}{48}},
  \bibinfo{pages}{1745} (\bibinfo{year}{1982}).

\bibitem[{\citenamefont{Gerber et~al.}(1997)\citenamefont{Gerber, Milner,
  Deutscher, Karpovsky, and Gladkikh}}]{gerber97}
\bibinfo{author}{\bibfnamefont{A.}~\bibnamefont{Gerber}},
  \bibinfo{author}{\bibfnamefont{A.}~\bibnamefont{Milner}},
  \bibinfo{author}{\bibfnamefont{G.}~\bibnamefont{Deutscher}},
  \bibinfo{author}{\bibfnamefont{M.}~\bibnamefont{Karpovsky}},
  \bibnamefont{and} \bibinfo{author}{\bibfnamefont{A.}~\bibnamefont{Gladkikh}},
  \bibinfo{journal}{Phys. Rev. Lett.} \textbf{\bibinfo{volume}{78}},
  \bibinfo{pages}{4277} (\bibinfo{year}{1997}).

\bibitem[{\citenamefont{Simon et~al.}(1987)\citenamefont{Simon, Dalrymple, {Van
  Vechten}, Fuller, and Wolf}}]{simon87}
\bibinfo{author}{\bibfnamefont{R.~W.} \bibnamefont{Simon}},
  \bibinfo{author}{\bibfnamefont{B.~J.} \bibnamefont{Dalrymple}},
  \bibinfo{author}{\bibfnamefont{D.}~\bibnamefont{{Van Vechten}}},
  \bibinfo{author}{\bibfnamefont{W.~W.} \bibnamefont{Fuller}},
  \bibnamefont{and} \bibinfo{author}{\bibfnamefont{S.~A.} \bibnamefont{Wolf}},
  \bibinfo{journal}{Phys. Rev. B} \textbf{\bibinfo{volume}{36}},
  \bibinfo{pages}{1962} (\bibinfo{year}{1987}).

\bibitem[{\citenamefont{Ando et~al.}(1995)\citenamefont{Ando, Boebinger,
  Passner, Kimura, and Kishio}}]{ando1}
\bibinfo{author}{\bibfnamefont{Y.}~\bibnamefont{Ando}},
  \bibinfo{author}{\bibfnamefont{G.~S.} \bibnamefont{Boebinger}},
  \bibinfo{author}{\bibfnamefont{A.}~\bibnamefont{Passner}},
  \bibinfo{author}{\bibfnamefont{T.}~\bibnamefont{Kimura}}, \bibnamefont{and}
  \bibinfo{author}{\bibfnamefont{K.}~\bibnamefont{Kishio}},
  \bibinfo{journal}{Phys. Rev. Lett.} \textbf{\bibinfo{volume}{75}},
  \bibinfo{pages}{4662} (\bibinfo{year}{1995}).

\bibitem[{\citenamefont{Boebinger et~al.}(1996)\citenamefont{Boebinger, Ando,
  Passner, Kimura, Okuya, Shimoyama, Kishio, Tamasaku, Ichikawa, and
  Uchida}}]{boebinger1}
\bibinfo{author}{\bibfnamefont{G.~S.} \bibnamefont{Boebinger}},
  \bibinfo{author}{\bibfnamefont{Y.}~\bibnamefont{Ando}},
  \bibinfo{author}{\bibfnamefont{A.}~\bibnamefont{Passner}},
  \bibinfo{author}{\bibfnamefont{T.}~\bibnamefont{Kimura}},
  \bibinfo{author}{\bibfnamefont{M.}~\bibnamefont{Okuya}},
  \bibinfo{author}{\bibfnamefont{J.}~\bibnamefont{Shimoyama}},
  \bibinfo{author}{\bibfnamefont{K.}~\bibnamefont{Kishio}},
  \bibinfo{author}{\bibfnamefont{K.}~\bibnamefont{Tamasaku}},
  \bibinfo{author}{\bibfnamefont{N.}~\bibnamefont{Ichikawa}}, \bibnamefont{and}
  \bibinfo{author}{\bibfnamefont{S.}~\bibnamefont{Uchida}},
  \bibinfo{journal}{Phys. Rev. Lett.} \textbf{\bibinfo{volume}{77}},
  \bibinfo{pages}{5417} (\bibinfo{year}{1996}).

\bibitem[{\citenamefont{Lang et~al.}(2002)\citenamefont{Lang, Madhavan,
  Hoffman, Hudson, Eisaki, Uchida, and Davis}}]{lang1}
\bibinfo{author}{\bibfnamefont{K.~M.} \bibnamefont{Lang}},
  \bibinfo{author}{\bibfnamefont{V.}~\bibnamefont{Madhavan}},
  \bibinfo{author}{\bibfnamefont{J.~E.} \bibnamefont{Hoffman}},
  \bibinfo{author}{\bibfnamefont{E.~W.} \bibnamefont{Hudson}},
  \bibinfo{author}{\bibfnamefont{E.}~\bibnamefont{Eisaki}},
  \bibinfo{author}{\bibfnamefont{S.}~\bibnamefont{Uchida}}, \bibnamefont{and}
  \bibinfo{author}{\bibfnamefont{J.~C.} \bibnamefont{Davis}},
  \bibinfo{journal}{Nature} \textbf{\bibinfo{volume}{415}},
  \bibinfo{pages}{412} (\bibinfo{year}{2002}).

\bibitem[{\citenamefont{Luttinger}(1964)}]{luttinger64}
\bibinfo{author}{\bibfnamefont{J.~M.} \bibnamefont{Luttinger}},
  \bibinfo{journal}{Phys. Rev.} \textbf{\bibinfo{volume}{135}},
  \bibinfo{pages}{A1505} (\bibinfo{year}{1964}).

\bibitem[{\citenamefont{Berkovits}(1999)}]{berkovits1999}
\bibinfo{author}{\bibfnamefont{R.}~\bibnamefont{Berkovits}},
  \bibinfo{journal}{Phys. Rev. \textbf{B}} \textbf{\bibinfo{volume}{60}},
  \bibinfo{pages}{26} (\bibinfo{year}{1999}).

\bibitem[{\citenamefont{Berkovits and Shklovskii}(1999)}]{berkshklovs1999}
\bibinfo{author}{\bibfnamefont{R.}~\bibnamefont{Berkovits}} \bibnamefont{and}
  \bibinfo{author}{\bibfnamefont{B.~I.} \bibnamefont{Shklovskii}},
  \bibinfo{journal}{J. Phys.: Condens. Mat.} \textbf{\bibinfo{volume}{11}},
  \bibinfo{pages}{779} (\bibinfo{year}{1999}).

\bibitem[{\citenamefont{Z\'arand et~al.}(2000)\citenamefont{Z\'arand,
  Zim\'anyi, and Wilhelm}}]{zarand2000}
\bibinfo{author}{\bibfnamefont{G.}~\bibnamefont{Z\'arand}},
  \bibinfo{author}{\bibfnamefont{G.~T.} \bibnamefont{Zim\'anyi}},
  \bibnamefont{and} \bibinfo{author}{\bibfnamefont{F.}~\bibnamefont{Wilhelm}},
  \bibinfo{journal}{Phys. Rev. \textbf{B}} \textbf{\bibinfo{volume}{62}},
  \bibinfo{pages}{8137} (\bibinfo{year}{2000}).

\bibitem[{\citenamefont{Efetov and Tschersich}(2002)}]{efetov02}
\bibinfo{author}{\bibfnamefont{K.~B.} \bibnamefont{Efetov}} \bibnamefont{and}
  \bibinfo{author}{\bibfnamefont{A.}~\bibnamefont{Tschersich}},
  \bibinfo{journal}{Europhysics Lett.} \textbf{\bibinfo{volume}{59}},
  \bibinfo{pages}{114} (\bibinfo{year}{2002}).

\bibitem[{\citenamefont{Loh et~al.}(2005)\citenamefont{Loh, Tripathi, and
  Turlakov}}]{lohgranular05}
\bibinfo{author}{\bibfnamefont{Y.~L.} \bibnamefont{Loh}},
  \bibinfo{author}{\bibfnamefont{V.}~\bibnamefont{Tripathi}}, \bibnamefont{and}
  \bibinfo{author}{\bibfnamefont{M.}~\bibnamefont{Turlakov}},
  \bibinfo{journal}{Phys. Rev. \textbf{B}}  \textbf{\bibinfo{volume}{72}},
  \bibinfo{pages}{233404} (\bibinfo{year}{2005}).

\bibitem[{\citenamefont{Shapira and Deutscher}(1984)}]{shapira84}
\bibinfo{author}{\bibfnamefont{Y.}~\bibnamefont{Shapira}} \bibnamefont{and}
  \bibinfo{author}{\bibfnamefont{G.}~\bibnamefont{Deutscher}},
  \bibinfo{journal}{Phys. Rev. \textbf{B}} \textbf{\bibinfo{volume}{30}},
  \bibinfo{pages}{166} (\bibinfo{year}{1984}).

\end{thebibliography}

\end{document}